\documentclass[intlimits,twoside,a4paper]{article}

\usepackage{amsmath,amssymb}
\usepackage{graphicx}

\usepackage[T2A]{fontenc}
\usepackage[cp1251]{inputenc}

\usepackage{color}

\usepackage[eqsecnum]{cmpj2}

\issue{2016}{19}{2}{23604}
\doinumber{10.5488/CMP.19.23604}

\title[Collective excitations in water]%
{Ab initio molecular dynamics study of collective excitations in liquid
H$_2$O and D$_2$O: \\ Effect of dispersion corrections}%
\author[T. Bryk, A.P. Seitsonen]{T. Bryk\refaddr{label1,label2},
        A.P. Seitsonen\refaddr{label3,label4}}
\addresses{
\addr{label1} Institute for Condensed Matter Physics of the National Academy of Sciences of Ukraine, \\
1 Svientsitskii St., 79011 Lviv, Ukraine
\addr{label2} Institute of Applied Mathematics and
Fundamental Sciences, Lviv Polytechnic National University, 79013
Lviv, Ukraine
\addr{label3}
Institut f\"{u}r Chemie, Universit\"{a}t Z\"{u}rich, Winterthurerstrasse 190, CH-8057 Z\"{u}rich, Switzerland
\addr{label4}
D\'{e}partement de Chimie, \'{E}cole Normale Sup\'{e}rieure, 24 rue Lhomond, F-75005 Paris, France}

\date{Received January 31, 2016, in final form February 12, 2016}
\authorcopyright{T. Bryk, A.P. Seitsonen, 2016}

\begin{document}
\maketitle

\begin{abstract}
The collective dynamics in liquid water is an active research topic experimentally, theoretically and via simulations.
Here, {\it ab initio} molecular dynamics simulations are reported in heavy and ordinary water
at temperature 323.15~K, or 50$^\circ$C. The simulations in heavy water were performed both with and without
dispersion corrections. We found that the dispersion correction (DFT-D3) changes
the relaxation of density-density time correlation functions from a slow, typical of
a supercooled state, to exponential decay behaviour of regular liquids. This implies an essential
reduction of the melting point of ice in simulations with DFT-D3. Analysis of longitudinal (L)
and transverse (T) current spectral functions allowed us to estimate the dispersions of acoustic
and optic collective excitations and to observe the L-T mixing effect. The dispersion
correction shifts the L and T optic (O) modes to lower frequencies and provides by almost thirty per cent smaller gap
between the longest-wavelength LO and TO excitations, which can be a consequence of
a larger effective high-frequency dielectric permittivity in simulations with dispersion corrections.
Simulation in ordinary water with the dispersion correction results in frequencies
of optic excitations higher than in D$_2$O, and in a long-wavelength
LO-TO gap of 24~ps$^{-1}$ (127~cm$^{-1}$).
\pacs 61.20.Ja, 61.20.Lc, 62.60.+v, 63.50.-x, 78.30.C-
\end{abstract}


\section {Introduction}

Collective dynamics in simple and molecular liquids is so far well understood only on macroscopic
length and time scales. Experiments on Brillouin scattering of light on liquids can be
very nicely described by the hydrodynamic theory, which explains the main relaxation and propagation processes
contributing to the measured scattered intensity of light~\cite{Han,Boo}.
However, hydrodynamic theory, which is a collection
of fundamental local conservation laws (balance equations) for a system treated as continuum, is
incapable of describing the atomic-scale dynamics and the  collective processes on nanoscale where the
atomistic structure of matter does not make it possible to treat the system as continuum. Atomistic molecular dynamics (MD)
computer simulations is a very efficient tool in exploration of the dynamic properties
of condensed matter and
provide a lot of precious information on time-dependent correlations in liquids on nano- and
atomic-scale resolution. However, the classical computer simulations make use of effective interparticle
potentials, which often do not permit, especially in the case of molecular liquids, to recover
the values of experimental melting points, $T_\text{m}$, or transport coefficients. In the case of water, the
difference between the calculated melting temperature from classical MD simulations and the
experimental melting point of ice I$_h$ can be over 80~K (with the SPC model)~\cite{Veg05}. Depending
on the values of effective ionic charges, massless charges used for correction of a dipole moment,
fixed intramolecular O--H bond length and the
H--O--H angle various water models result in different melting points which are usually lower than the
experimental value~\cite{Veg05}.

{\it Ab initio} molecular dynamics (AIMD) simulations, which use the density functional theory (DFT) for
an electronic subsystem and
classical equations of motion for ions, are free of effective interatomic potentials because
the input interactions for simulations are electron-ion pseudopotentials. On-the-fly minimization of total energy to the Born-Oppenheimer surface
and the consequent equilibrium AIMD runs make it possible to study instantaneous distributions of electron density that
contain all the information about the bonding and polarization fluctuations in a system.
The AIMD of water and ice also face some problems from the point of view of the correct melting point of
ice I$_h$. Recently, two-phase ice/water AIMD simulations with Becke-Lee-Yang-Parr
(BLYP)~\cite{Becke_PRA_1988a,Lee_PRB_1988a}
exchange-correlation functional indicated that at ambient pressure, the melting point of ice I$_h$
should be more than 400~K, but taking into account the so-called (London) dispersion corrections to BLYP,
the melting temperature was lowered to about 360~K~\cite{Yoo11}.
Namely, one of the problems of DFT calculations with generalized gradient approximations (GGA), such as BLYP, is that DFT-GGA does not reproduce the long-range attractive
interaction (typical of Lennard-Jones effective potentials), whose leading term in atom pairs is $\propto -r^{-6}$.
The semiempirical dispersion correction proposed by Grimme~\cite{Gri04} to account for van der
Waals interactions in DFT-GGA calculations resulted in good agreement with experimental results
in binding energies and intermolecular distances for a variety of molecular complexes.
Another approach to account for van~der~Waals corrections in DFT calculations was initially proposed by Dion {et al.}
\cite{Dio04}. Such corrections were used to study the effect of dispersion correction on the structure and
self-diffusion in water \cite{Wan11}.

One can also mention  another direction in attempts to improve
static and dynamic properties of water from AIMD through application of different hybrid density
functionals by making use of short-ranged Hartree-Fock exchange \cite{Tod06}, or the recent truly \textit{ab initio} MD simulations using the MP2 perturbation theory for correlations beyond the Hartree-Fock energy \cite{DelBen_2015_a}.
A newer generation (D3) of the dispersion corrections was suggested in 2010~\cite{Grimme_JCP_2010a}. No AIMD simulations have been reported so far to check how the D3 dispersion correction would affect the
value of the calculated melting point of ice I$_h$.

Furthermore, no analysis yet regarding the time correlation functions and regarding the collective dynamics of water from AIMD studies in which dispersion corrections would be used in simulations. The collective
dynamics of water itself is not completely understood, although many features such as the
 ``fast sound''
phenomenon \cite{Rah74,Ruo08}, rotational and structural relaxation in water~\cite{Ruo99},
and the gap existing between the
longitudinal and transverse optic modes have been known for years. To date, there has been
no generalized hydrodynamic approach to the collective dynamics in water capable of  recovering the simulation
results with full account for rotational and structural relaxations, although some successful
attempts were made mainly to describe polarization and
dielectric relaxation in water~\cite{Bop98,Ome00,Sed14,Elt16}. As concerns heavy water D$_2$O, only
a few studies were devoted to the analysis of collective dynamics \cite{Sam97,She02,Sac04}.
In reference~\cite{Sac04}, the
findings for heavy water were compared to the previous experimental scattering studies on H$_2$O and
the  existence of a very flat dispersion curve at $\sim 6$~meV was attributed to optic modes.

A very interesting finding for heavy water was reported from  classical MD simulations using SPC/E model \cite{Sam97}: the
analysis of longitudinal (L) and transverse (T) current spectral functions $C^\textrm{L/T}(k,\omega)$,
calculated directly
from the trajectories and the corresponding velocities of particles, showed a striking feature which later on was referred to as a mixing between L and T dynamics in water: Starting from wave numbers $\sim 0.5$~{\AA}$^{-1}$,
the L- and T-current spectral functions showed the emergence of a shoulder at a frequency
corresponding to the T- and L-excitations, which become well defined
peaks at higher wave numbers. These peaks gave evidence of both (L and T) excitations contributing to L- and T-current spectral
functions. Later on, the inelastic X-ray scattering (IXS) experiments on water~\cite{Cim10} completely
supported the MD findings on the observation of the L-T coupling in reference~\cite{Sam97}. Although there is
a clear hint that rotational motion of molecules can cause the L-T coupling, so far there has not
appeared either any derivation of a generalized hydrodynamic theory for water with coupled equations for L and T dynamics
capable of explaining the MD or experimental observations on L-T mixing in water. We can mention
here that the issue of the L-T coupling is not solely restricted to water or molecular liquids
in general: The L-T coupling effects were also observed while analysing the IXS scattering experiments
on liquid Hg~\cite{Hos09} and Sn~\cite{Hos13}, as well as reported from the observation of
a two-peak structure of current spectral functions obtained in {\it ab initio} simulations of
liquid Sn~\cite{Mun12} and liquid Li at high pressures \cite{Bry15}. No L-T coupling
effects were studied so far in water by {\it ab initio} simulations.
Recently, there was a report on classical simulations
of supercooled water at temperature 180~K using TIP4P model~\cite{Jed11}. It was found from a fit of
several Lorentzians that for
supercooled water, transverse current spectral functions contained contributions from four collective
modes, while for longitudinal dynamics, only three contributions from collective excitations were
estimated. Furthermore, it was shown that for the three studied systems (with density 0.939~g/cm$^3$,
1.029~g/cm$^3$ and 1.179~g/cm$^3$) the high-frequency speed of longitudinal acoustic modes increased
from 3313~m/s to 3679~m/s.

Very recently, a classical simulation study of high-frequency dynamics of water using two
TIP4P/2005f (with 512 molecules) and TTM3F (with 256 molecules) water models was
reported~\cite{Elt16}. The main focus was on calculations
of polarization time correlation functions and estimation of dispersion of L and T optic (LO and TO, respectively) modes.
The gap between the longest-wavelength LO and TO excitations obtained at the smallest accessible
wave numbers in those simulations was larger than the experimental values of LO-TO gap at the
corresponding temperatures. One of the shortcomings of classical MD simulations for ionic and
polar liquids with non-polarizable and even polarizable effective potentials is the neglect (for
non-polarizable models) or a modelling (for polarizable models) of high-frequency polarizability of
electron subsystem. It was shown in reference~\cite{Bry08}, from a comparison of classical simulations using
a rigid ion model for molten salt NaCl and {\it ab initio} simulations at the same thermodynamic
point, that the high-frequency polarizability leads to a ``softening'' of the long-wavelength LO
frequency by a factor $\propto \sqrt{\varepsilon_{\infty}}$, while the transverse TO frequency becomes
slightly higher (TO ``hardening'') in comparison with the non-polarizable case. Here,
$\varepsilon_{\infty}$ is the high-frequency dielectric permittivity. These findings were
supported by sequential AIMD simulations for molten salts NaI~\cite{Bry09},
RbF~\cite{Bry10}
and LiBr~\cite{Bry12}. Therefore, it would be interesting to study the LO-TO gap for most
long-wavelength excitations in water from {\it ab initio} simulations.

In this study, we aimed at calculating, from AIMD, the L- and T-current spectral functions and
estimating the dispersion of collective excitations. Initially, we intended to simulate only
heavy water D$_2$O both with and without dispersion corrections to DFT in order to estimate the
effect of dispersion corrections on the decay of density-density time correlation functions,
which is very sensitive to the regular liquid/supercooled/glassy state of the studied system,
as well as to observe the effect of dispersion corrections on the high-frequency speed of sound
and LO, TO frequencies. Additional simulations were performed in light, or ``normal'' water, with dispersion
corrections applied, which enabled us to compare our results on the dispersion of LO and TO
excitations with the recently reported ones~\cite{Elt16}. Another important issue was to estimate
whether the L-T coupling effects are observed in {\it ab initio} simulations of water and
heavy water, and to estimate in general in what exactly consists the difference in dispersion of collective excitations
between heavy and regular water.

The remaining Paper is organized as follows: In the next
section we supply details of our {\it ab initio} simulations, while the third section contains
results for static structure of D$_2$O obtained from simulations with and without dispersion
corrections; the static structure in light water is practically identical to that in heavy water. Further, we will
show the results for the density-density time correlation functions, L- and T-current spectral functions,
and dispersions of collective excitations. The last section contains conclusions of this study.

\section{Molecular dynamics simulations}

We performed {\it ab initio} simulations having a collection of 128 molecules
in a periodic box with the box length of 15.7459~{\AA} for three systems: heavy water
D$_2$O with D3 dispersion correction and without it, and ordinary water H$_2$O
with D3 dispersion correction.
The details of our simulations are very similar to those used in an earlier
work~\cite{Jonchiere_JCP_2011a}. We performed Born-Oppenheimer molecular
dynamics simulations using the density functional theory to provide the
forces into the Verlet algorithm. We integrated the equations of motion with a
time step of 0.5~fs in the canonical, $NVT$, ensemble, and used a
Nos\'{e}-Hoover thermostat chain with a time constant of 2.4~ps to equilibrate
the average temperature at 323.15~K. The slightly elevated temperature is
motivated by the known deficiency of the approximations in the details of the
DFT treatment discussed
below \cite{Lin_JPCB_2009a,Schmidt_JCP_2009a,Jonchiere_JCP_2011a}.
Either the mass of hydrogen or deuterium was used in
simulations of light and heavy water, respectively. All the trajectories are
at least 50~ps long, and the first 10 ps were neglected as a period of
equilibration.

We used the BLYP generalized gradient
approximation as the exchange-correlation term in the Kohn-Sham equations of
DFT. Further, the D3 approximation~\cite{Grimme_JCP_2010a} was added
 to the Kohn-Sham expression of total energy in order to
incorporate the London dispersion forces missing in the BLYP, or (semi-)local approximations in general.

The CP2K suite of programs~\cite{CP2K} was used in the simulations,  in
particular its QuickStep~\cite{vandeVondele_CPC_2005a} module. The
Gaussian Plane Wave (GPW) method~\cite{Lippert_MolPhys_1997a} was used with
the triple-zeta valence doubly polarized (TZV2P) basis set of Gaussians and
plane waves up to 400~Ry to expand the Kohn-Sham orbitals and the electron
density, respectively. We also employed the NN50 smoothing
algorithm~\cite{vandeVondele_CPC_2005a} to reduce the errors in the evaluation
of the energy and forces of the BLYP term, since tests~\cite{Jonchiere_JCP_2011a}
have shown that this combination delivers a good convergence in the
simulations. The action of the nuclei and core electrons on the valence
electrons was replaced by the Goedecker-Teter-Hutter (GTH) norm-conserving
pseudo-potentials \cite{Goedecker_PRB_1996a,Hartwigsen_PRB_1998a}.

At each configuration, we calculated Fourier-components of the following
dynamic variables: partial particle densities
\begin{equation} \label{nkt}
n_{\alpha}(k,t)=\frac{1}{\sqrt{N_{\alpha}}}\sum_{i=1}^{N_{\alpha}}
\re^{-\ri{\bf k}{\bf r}^{\alpha}_i(t)}\;, \qquad \alpha=\textrm{H/D, O}
\end{equation}
densities of partial longitudinal momentum
\begin{equation} \label{jlkt}
J^\text{L}_{\alpha}(k,t)=\frac{m_{\alpha}}{k\sqrt{N_{\alpha}}}\sum_{i=1}^{N_{\alpha}}
({\bf k}{\bf v}^{\alpha}_i)
\re^{-\ri{\bf k}{\bf r}^{\alpha}_i(t)}\;, \qquad \alpha=\textrm{H/D, O}
\end{equation}
and densities of partial transverse momentum
\begin{equation} \label{jtkt}
{\bf J}^\text{T}_{\alpha}(k,t)=\frac{m_{\alpha}}{k\sqrt{N_{\alpha}}}\sum_{i=1}^{N_{\alpha}}
[{\bf k}{\bf v}^{\alpha}_i]
\re^{-\ri{\bf k}{\bf r}^{\alpha}_i(t)}\;, \qquad \alpha=\textrm{H/D, O}.
\end{equation}
Here, $N_\textrm{O}$ and $N_\textrm{H/D}$ are 128 and 256, respectively, in our simulations.
The smallest accessible wave number from our AIMD was 0.399~\AA$^{-1}$.
We studied the range of wave numbers up to 4.5~\AA$^{-1}$.
All the wave number-dependent quantities calculated from AIMD were averaged over all possible
directions of wave vectors corresponding to the same absolute value.
The partial
densities were used for calculations of collective Bhatia-Thornton time
correlation functions and corresponding static structure factors either in ``T--C'' (total
number--concentration) or in ``M--X'' (total mass--mass-concentration) representations~\cite{Bha74}.
All these partial and collective representations contain the full information about the same
collective mode, although, as it was shown in \cite{Bry00a,Bry00b}, the mass-concentration
currents are the most useful ones in the studies of optic collective modes.

\section{Results and discussion}

\begin{figure}[!t]
\centerline{
\includegraphics[width=0.95\textwidth]{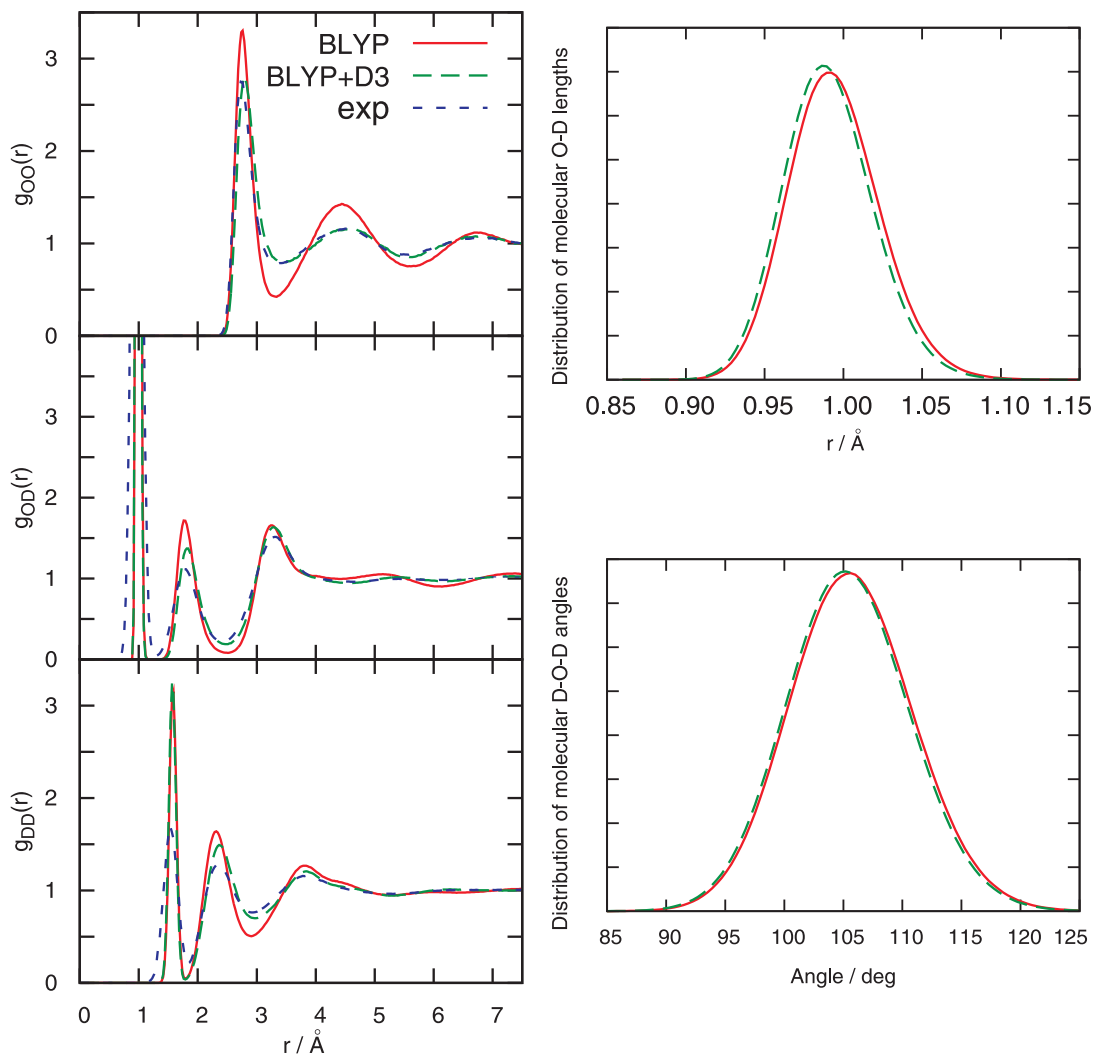}%
}
\caption{(Color online) Partial pair distribution functions $g_{ij}(r)$
and distributions of intramolecular O--D lengths and D--O--D angles for
D$_2$O at $T=323.15$~K obtained from BLYP (red solid line) and BLYP+D3
(dashed green line) simulations. The amplitude of intramolecular peak
in $g_\textrm{OD}(r)$ is 33.6 and 34.6 with BLYP and BLYP+D3, respectively.
Partial pair distribution functions and distributions of
intramolecular bond lengths and angles in H$_2$O with BLYP+D3 were
practically identical to D$_2$O-BLYP+D3 results. The experimental data
for water at $T=298$~K and pressure 1~bar were taken from reference \cite{Sop00}.
} \label{rdf}
\end{figure}

Partial pair distribution functions from simulations with and without the D3 dispersion correction
are shown in figure~\ref{rdf}. The simulations of D$_2$O and H$_2$O both with dispersion
correction resulted in practically identical static, or average, structure. One can see in figure~\ref{rdf} that the
D3 correction led to a strong reduction in the depth of the first minimum of partial O--O distribution compared to the
regular BLYP simulations, which resulted in an overbonded water molecules; this is
related to a very high melting temperature of ice I$_h$ from AIMD with BLYP exchange-correlation
functional, suggested in reference~\cite{Yoo11}. The height of all the first intermolecular peaks in the partial distribution functions
was reduced in BLYP+D3 simulations, and first intermolecular minima shifted to larger values.
This means that there is more active exchange of molecules in the first coordination shell with molecules
from the bulk, being consistent with an enhanced diffusion constant found in reference~\cite{Jonchiere_JCP_2011a}. The inclusion
of the D3 dispersion correction slightly reduced the values of intramolecular
O--H distance and H--O--H angle, as in the right-hand frames of figure~\ref{rdf} one can see that the
corresponding distributions were shifted towards smaller values in simulations with BLYP+D3.
In simulations with classical equations of motion for ions, there is practically no difference
in the results obtained for H$_2$O and D$_2$O, which appears in quantum treatment via path integral
 simulations~\cite{Che03,Sop08}. In figure~\ref{rdf}, by blue short-dash line we show the experimental
data for partial distribution functions in water at $T=298$~K \cite{Sop00}, which are in agreement
with our BLYP+D3 simulations, especially this being clearly seen in the O--O pair distribution
function.

\begin{figure}[!t]
\centerline{
\includegraphics[width=0.5\textwidth]{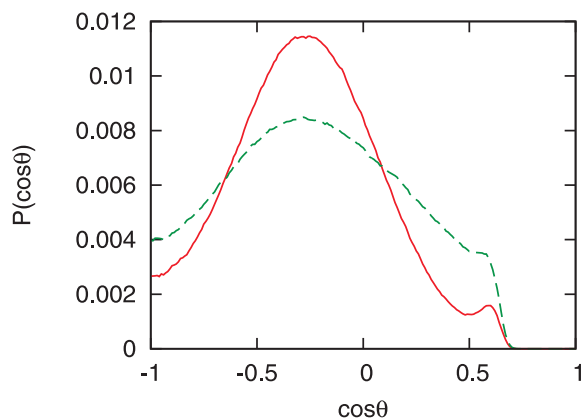}%
}
\caption{(Color online) Angular O--O--O distribution functions in
D$_2$O at $T=323.15$~K from BLYP (red solid line) and BLYP+D3
(dashed green line) {simulations, obtained within the radial cut-off of
3.375~{\AA}}. Results in H$_2$O with BLYP+D3 were
practically identical to D$_2$O ones with BLYP+D3. Path integral simulations
analyzed in reference \cite{Sop08} yield in H$_2$O and D$_2$O results in
reasonable agreement with our BLYP+D3 simulations.
} \label{bonddf}
\end{figure}

The angular O--O--O distribution functions in the first coordination shell
are shown in figure~\ref{bonddf}. One can compare these distributions with similar
ones reported from classical simulations of SPC/E water~\cite{Bry04,Gal15}.
There is a resemblance in the angular O--O--O distributions from BLYP simulations
with the results from SPC/E model in the temperature range 200--220~K
that is in an supercooled
state, because the estimates for the melting point of SPC/E model from
two-phase classical simulations resulted in 225~K~\cite{Bry02,Bry04};
thermodynamic integration technique for SPC/E model resulted in a somewhat
lower value of 215~K~\cite{Veg05}. The large peak at about $105^{\circ}$
corresponds to tetrahedral ordering of the nearest neighbors. One can see that
the application of D3 dispersion correction significantly reduces the contribution
from tetrahedral configurations of nearest neighbors, which is actually in line with
a suggestion of the reduction of the melting point of ice in simulations
with D3 dispersion correction. In the literature one can find a similar analysis
of O--O--O angle distributions for H$_2$O and D$_2$O, obtained from path integral simulations
\cite{Sop08}. Both O--O--O angle distributions in \cite{Sop08} show pronounced
maxima at  about $100^{\circ}-105^{\circ}$ and a small shoulder close to $60^{\circ}$,
which is in agreement with our BLYP+D3 simulations.

\begin{figure}[!b]
\centerline{
\includegraphics[width=0.5\textwidth]{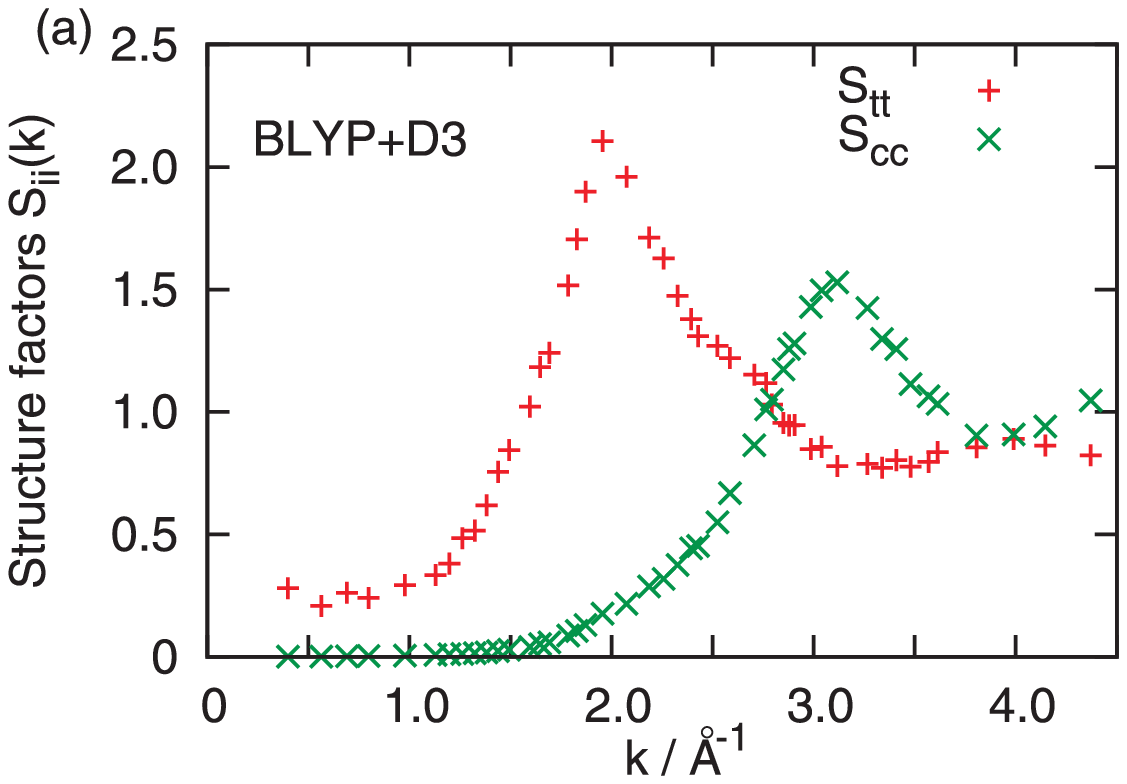}
\includegraphics[width=0.48\textwidth]{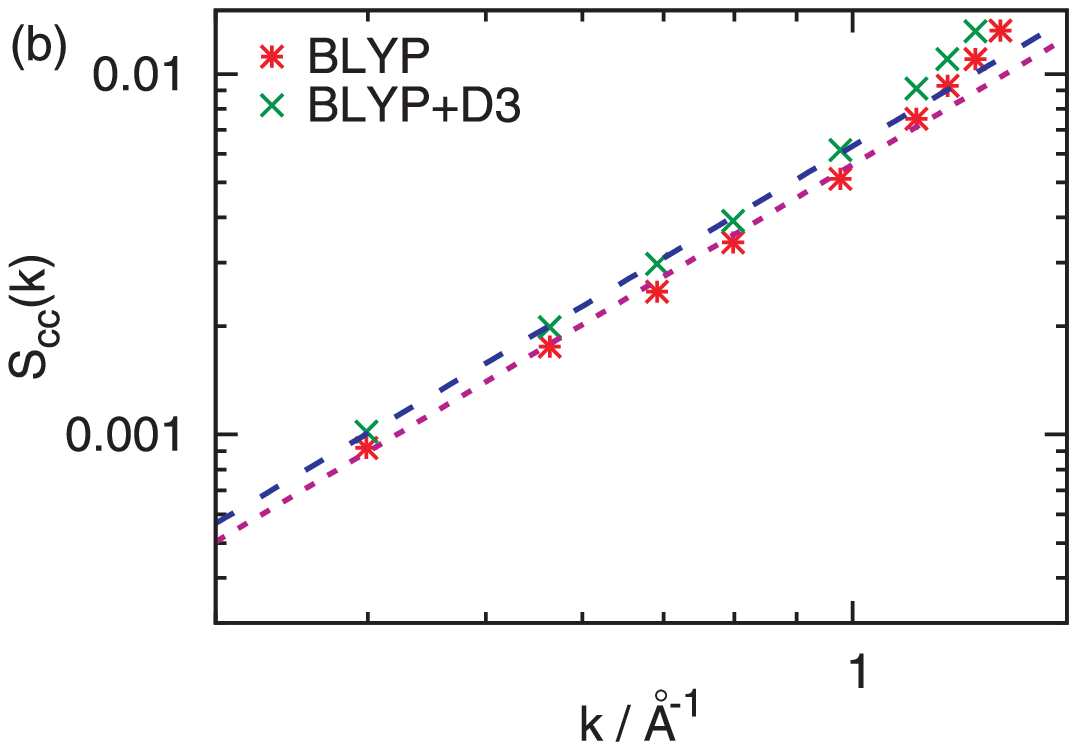}
}
\caption{(Color online) Bhatia-Thornton static structure factors in
H$_2$O at $T=323.15$~K in BLYP+D3 approximation (a)
and comparison of the long-wavelength asymptotes of the concentration
structure factors from BLYP and BLYP+D3
{simulations of D$_2$O (b)}. The straight lines in (b) correspond to a fit with $k^2$
dependence.
\label{strfact} }
\end{figure}

Partial static structure factors are of huge importance in understanding the role
of partial contributions to the measured X-ray or neutron diffraction intensities.
Here, we report the Bhatia-Thornton total density $S_\text{tt}(k)$ and concentration
$S_\text{cc}(k)$ static structure factors. The total density structure factor shown in
figure~\ref{strfact} has the main peak at $\sim 2$~{\AA}$^{-1}$ and a well-defined shoulder
in wave number range  $k\sim 2.5-2.8$~{\AA}$^{-1}$, which is in agreement with experimental
results~\cite{Hur00} and {\it ab initio} calculations of X-ray spectra of liquid water
\cite{Kra02}. The concentration structure factor $S_\text{cc}(k)$ has a well-defined
maximum at $k\sim 3.1$~{\AA}, which is related to short-range order of hydrogen bonds.
The concentration structure factor should have the same long-wavelength asymptote
of $\sim k^2$  as the charge structure factor. In figure~\ref{strfact}~(b), we show the
long-wavelength asymptotes of concentration structure factors in simulations with and
without D3 dispersion correction, which nicely recover the $\sim k^2$ behaviour.
The coefficient at the $k^2$ term depends on the value of the high-frequency dielectric
permittivity $\varepsilon_{\infty}$, which makes it possible to estimate $\varepsilon_{\infty}$
if classical simulations with non-polarizable effective interaction models were
available at the same thermodynamic point, like it was done in reference~\cite{Bry08} in liquid NaCl. Here, figure~\ref{strfact}~(b) implies that the system simulated with D3 dispersion correction has a larger value of (effective) $\varepsilon_{\infty}$. This fact is important in analysing the frequencies
of longitudinal optic modes that experience a softening for a system with larger values
of $\varepsilon_{\infty}$. Electron structure simulations further permit to perform calculations of more
general total charge structure factors $S_{QQ}(k)$, where $Q(k,t)$ are the total charge
fluctuations composed of point positive ions and instantaneous distributions of electron density
in AIMD.
Recently, it was shown that the long-wavelength asymptote of $S_{QQ}(k)$ in water
recovers the $\sim k^2$ behaviour~\cite{Kle14}.

\begin{figure}[!t]
\centerline{
\includegraphics[width=0.5\textwidth]{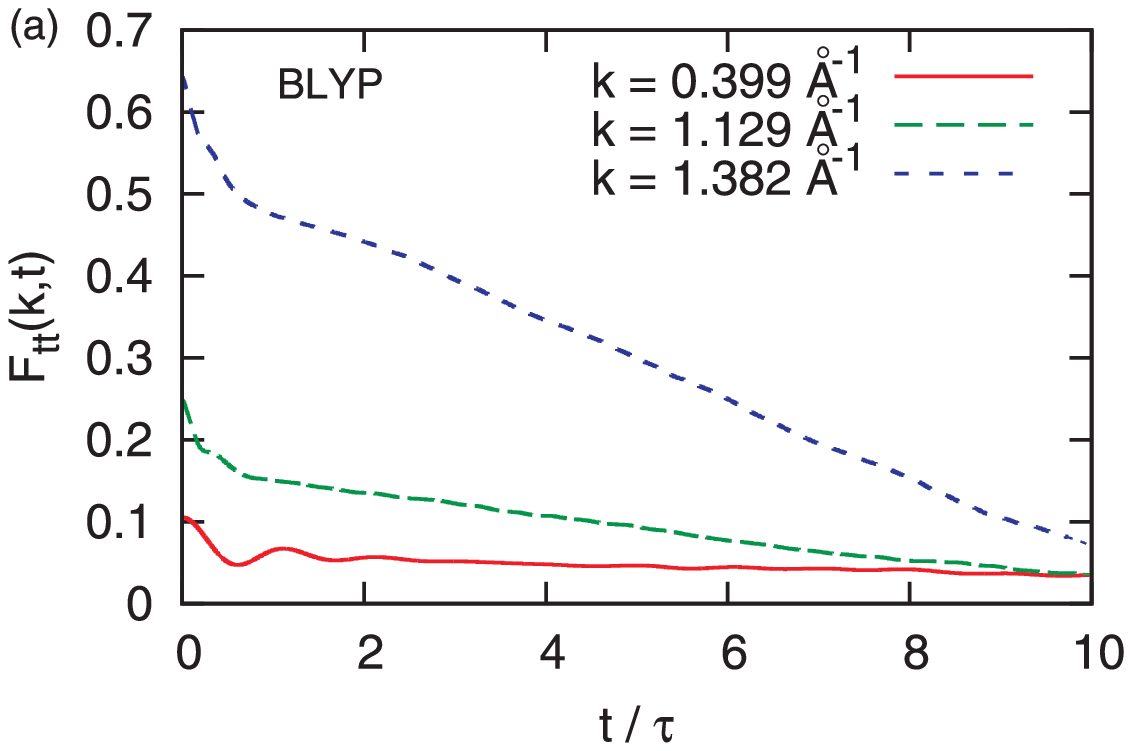}
\includegraphics[width=0.5\textwidth]{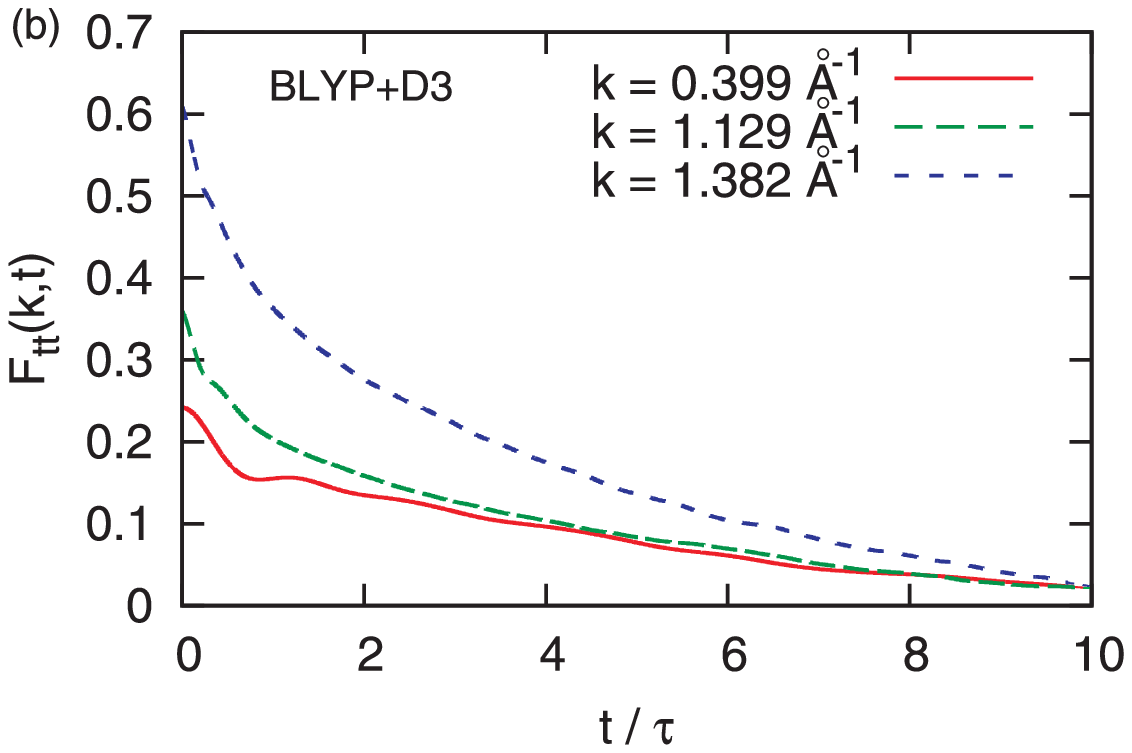}
}
\caption{(Color online) Total density autocorrelation functions for
three wave numbers from BLYP (a) and BLYP+D3 (b) simulations of D2O at $T = 323.15$~K.
The time unit $\tau=0.39475$~ps.
The $F_\text{tt}(k,t)$ in (a) contain
long tails which are typical of slow stretched-exponential relaxation
in undercooled liquids, while in (b) the relaxation of $F_\text{tt}(k,t)$
is typically a simple exponential one as for regular liquids.
} \label{ftt}
\end{figure}

The total density-density time correlation functions $F_\text{tt}(k,t)$ at three
wave numbers, calculated from simulations of D$_2$O with and without D3 dispersion
correction, are shown in figure~\ref{ftt}. In general, the decay of density-density time
correlation functions is very sensitive to the studied thermodynamic point on the
phase diagram. At low wave numbers, the hydrodynamic theory predicts leading contributions
to $F_\text{tt}(k,t)$ coming from longitudinal collective excitations (and represented by
a damped oscillator) and from thermal relaxation having an exponential tail.
The exponential decay of the tail of density-density time correlation inherent to
an ordinary liquid state is replaced by a slow stretched-exponential decay
if the liquid becomes supercooled. By approaching the glass transition,
the tail of the $F_\text{tt}(k,t)$ shows features connected with $\alpha$-relaxation, and ultimately,
at the glass transition, there appears a non-decaying plateau $F_\text{tt}(k,t\to \infty)=f(k)$
known as the non-ergodicity factor \cite{Goe00}. Moreover, for supercritical fluids it was recently shown
that the analysis of density-density correlations allows one to discriminate between
liquid-like and gas-like types of fluid in the supercritical regime \cite{Sim10,Bry10b}.

In figure~\ref{ftt}, one can see that the total density time correlation functions obtained
from simulations without D3 dispersion correction show
slow relaxations typical of supercooled liquids, while our AIMD simulations with D3 dispersion correction give evidence of regular exponential tails of $F_\text{tt}(k,t)$. This is a further proof that the simulated
thermodynamic point is below the melting line in simulations without accounting for dispersion
corrections, while BLYP+D3 simulations result in density-density correlations typical
of a regular liquid state.

\begin{figure}[!h]
\centerline{
\includegraphics[width=0.498\textwidth]{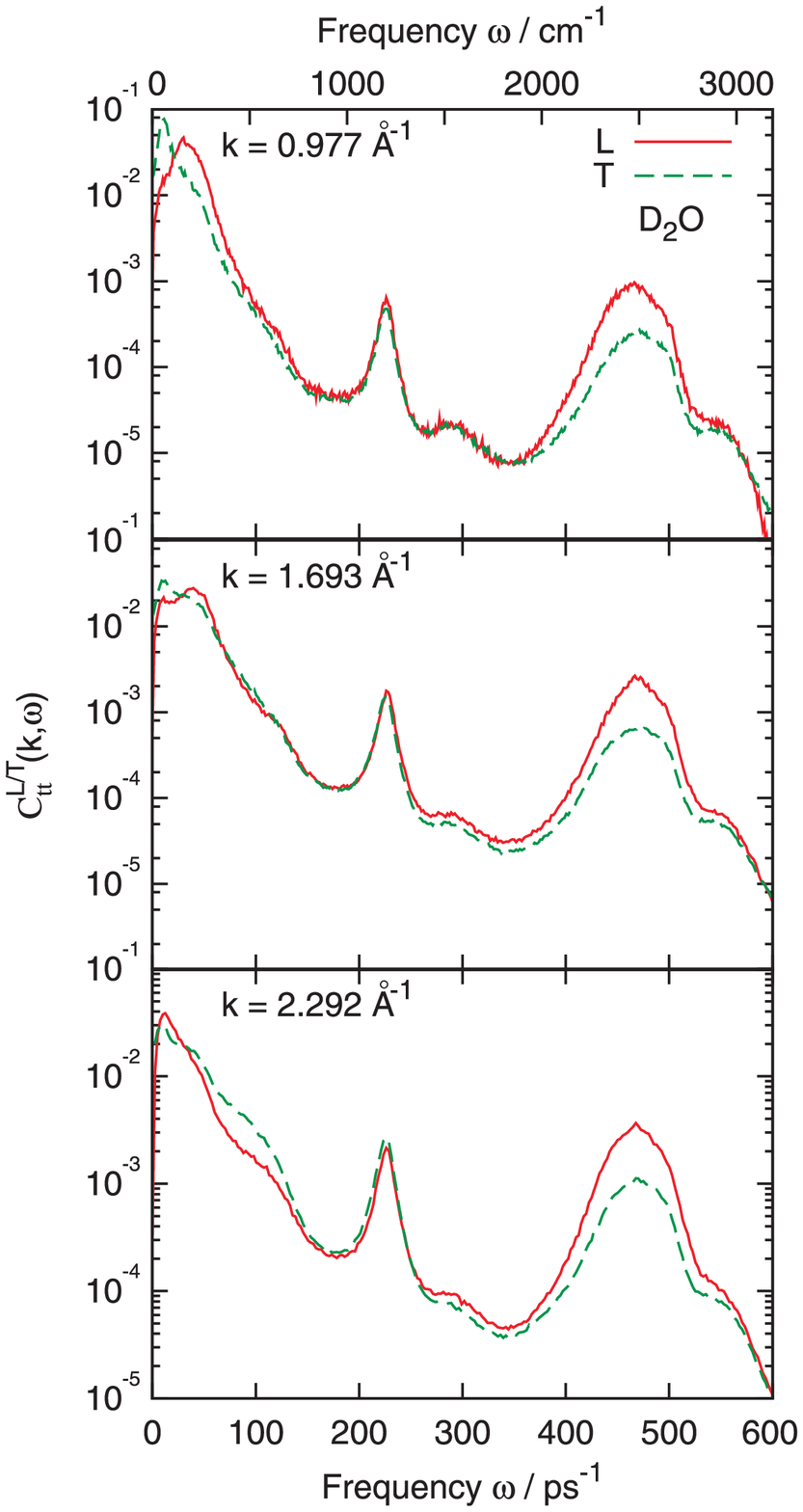} \
\includegraphics[width=0.48\textwidth]{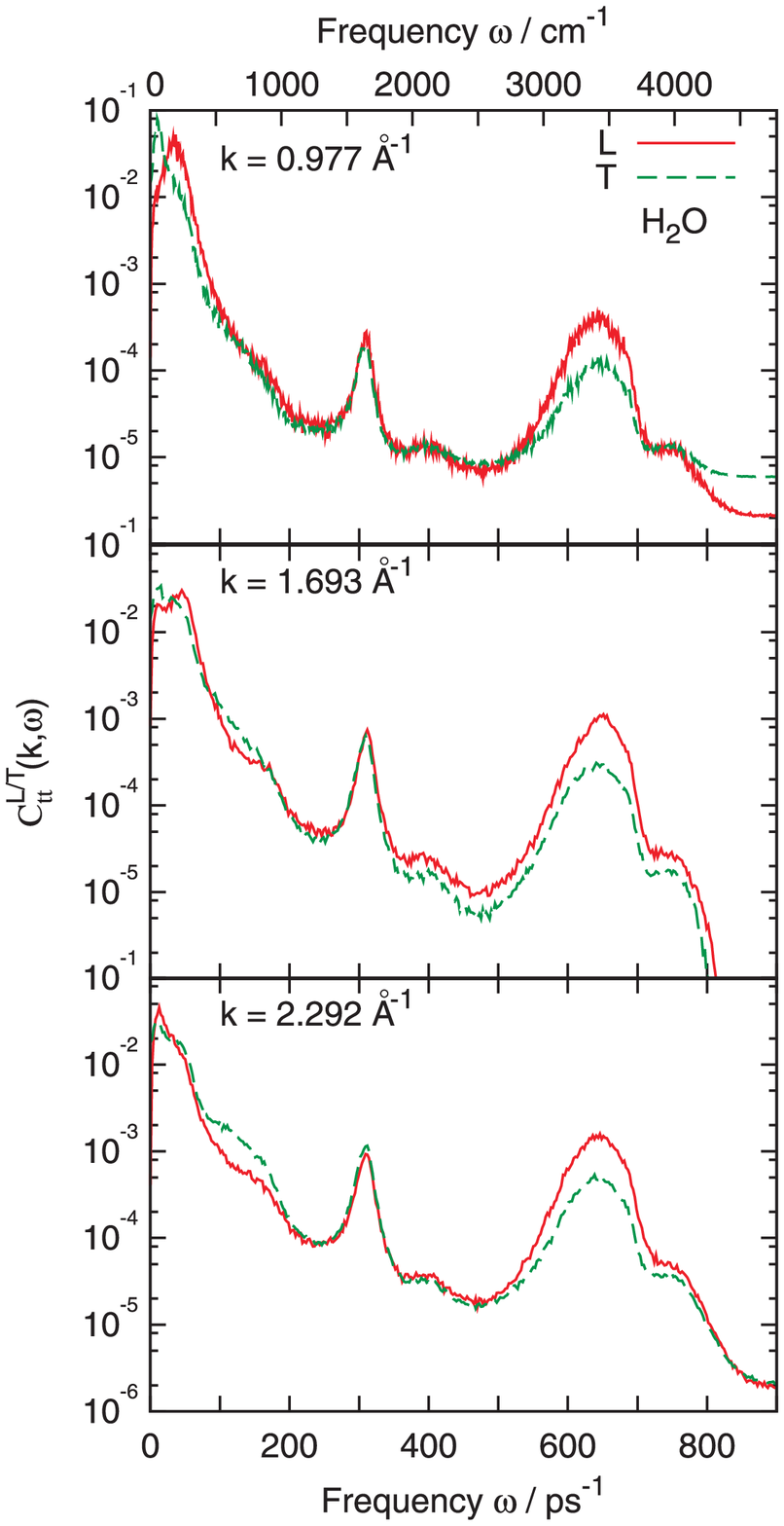}
}
\vspace{-3mm}
\parbox[t]{0.5\textwidth}{
\caption{(Color online)  Total longitudinal (L) and transverse (T) current spectral function
$C^\textrm{L/T}_\text{tt}(k,\omega)$ at
three wave numbers from BLYP+D3 simulations of D$_2$O at $T = 323.15$~K. \label{cttD2O}}
}
\parbox[t]{0.5\textwidth}{
\caption{(Color online) Total longitudinal (L) and transverse (T) current spectral function
$C^\textrm{L/T}_\text{tt}(k,\omega)$ at
three wave numbers from BLYP+D3 simulations of H$_2$O at $T = 323.15$~K.\label{ctt} }
}
\end{figure}

Longitudinal and transverse total current spectral functions $C^\textrm{L/T}_\text{tt}(k,\omega)$
obtained from simulations of D$_2$O (figure~\ref{cttD2O}) and H$_2$O (figure~\ref{ctt})
with D3 dispersion correction contain contributions from propagating L and T collective
excitations as well as from intramolecular normal modes. Due to a larger mass of D with respect
to H, the intramolecular part in D$_2$O is located in the range of frequencies 180--600~ps$^{-1}$,
while in light water, the intramolecular vibrations are at higher frequencies, 300--800~ps$^{-1}$. The
L and T total current spectral functions have a high-frequency peak corresponding to
dispersionless (localized) vibrations
at frequencies $\sim 226$~ps$^{-1}$ in D$_2$O (figure~\ref{cttD2O}) and $\sim 310$~ps$^{-1}$ in
H$_2$O (figure~\ref{ctt}). Simulations with and without D3 dispersion correction in D$_2$O
resulted practically in the same frequency.
{These peaks correspond to intramolecular bending normal modes, which have nearly two times lower frequencies
than those of the symmetric stretch normal modes.}

The frequencies of L and T acoustic collective excitations can be observed as the low-frequency
peaks in the corresponding total current spectral functions $C^\textrm{L/T}_\text{tt}(k,\omega)$. In figures~\ref{cttD2O} and \ref{ctt} one can see the emergence of another low-frequency peak at $k>1$~{\AA}$^{-1}$
in $C^{L}_\text{tt}(k,\omega)$, which practically coincides with the frequency of the transverse excitation.
This is the effect of L-T mixing reported previously from classical MD simulations~\cite{Sam97} and
X-ray scattering experiments \cite{Cim10}.

\begin{figure}[!t]
\centerline{
\includegraphics[width=0.5\textwidth]{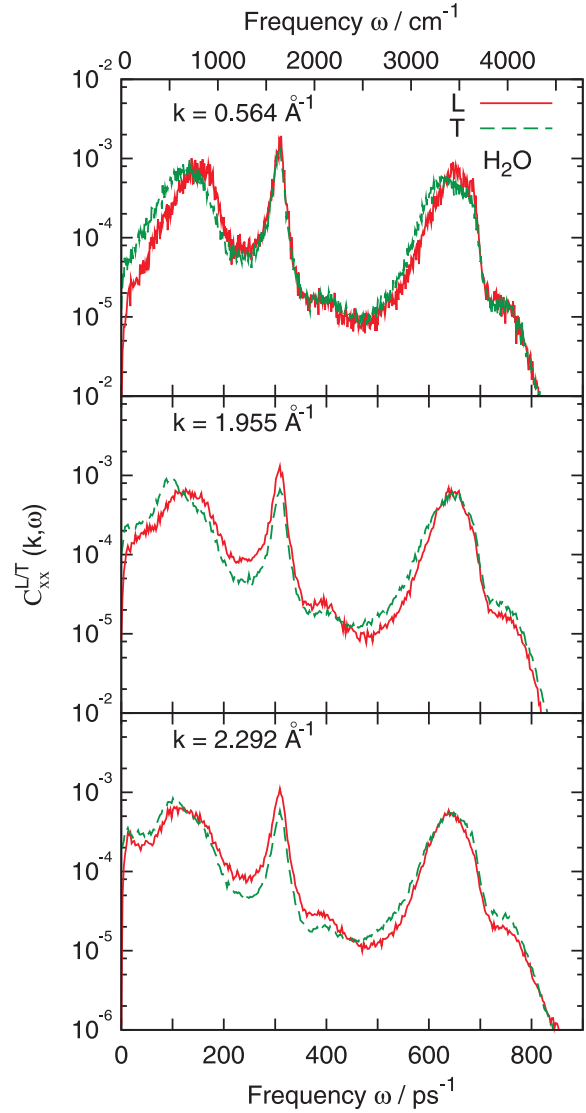}
}
\vspace{-2mm}
\caption{(Color online)  Mass-concentration current spectral function $C^\textrm{L/T}_{xx}(k,\omega)$ at
three wave numbers from BLYP+D3 simulations of H$_2$O at $T = 323.15$~K.
} \label{cxx}
\end{figure}

One can easily observe the optic-like modes  as peaks in the mass-concentration current spectral
functions $C^\textrm{L/T}_{xx}(k,\omega)$, shown for light water in figure~\ref{cxx}. A nice property of
$C^\textrm{L/T}_{xx}(k,\omega)$ spectral function is that  the contributions from acoustic modes in it
are suppressed and the high-frequency optic modes are well observable, in contrast to the
$C^\textrm{L/T}_\text{tt}(k,\omega)$ spectral function. The L and T optic modes clearly appear
at different frequencies in figure~\ref{cxx}, that is typical for ionic melts and polar fluids, and result in a LO-TO gap in the
long-wavelength region. For ionic crystals, there is a famous Lyddane-Sachs-Teller
relation for the difference between long-wavelength LO and TO phonons (non-damped propagating modes
of small displacements of positive/negative ions from equilibrium positions with opposite
phases) \cite{Lyd41}
\begin{equation} \label{LST}
\frac{\omega^2_\text{LO}}{\omega^2_\text{TO}}=\frac{\varepsilon(0)}{\varepsilon_{\infty}}~,
\end{equation}
where $\varepsilon(0)$ is the macroscopic static dielectric permittivity of a non-conducting system. Although in non-conducting liquids there should be some corrections to the Lyddane-Sachs-Teller
due to the damping of optic modes, which in its turn renormalizes the observed LO and TO frequencies,
this relation gives a hint of how the high-frequency screening by the electron density reduces the LO-TO
gap.

\begin{figure}[!t]
\begin{center}
\quad
\includegraphics[width=0.65\textwidth]{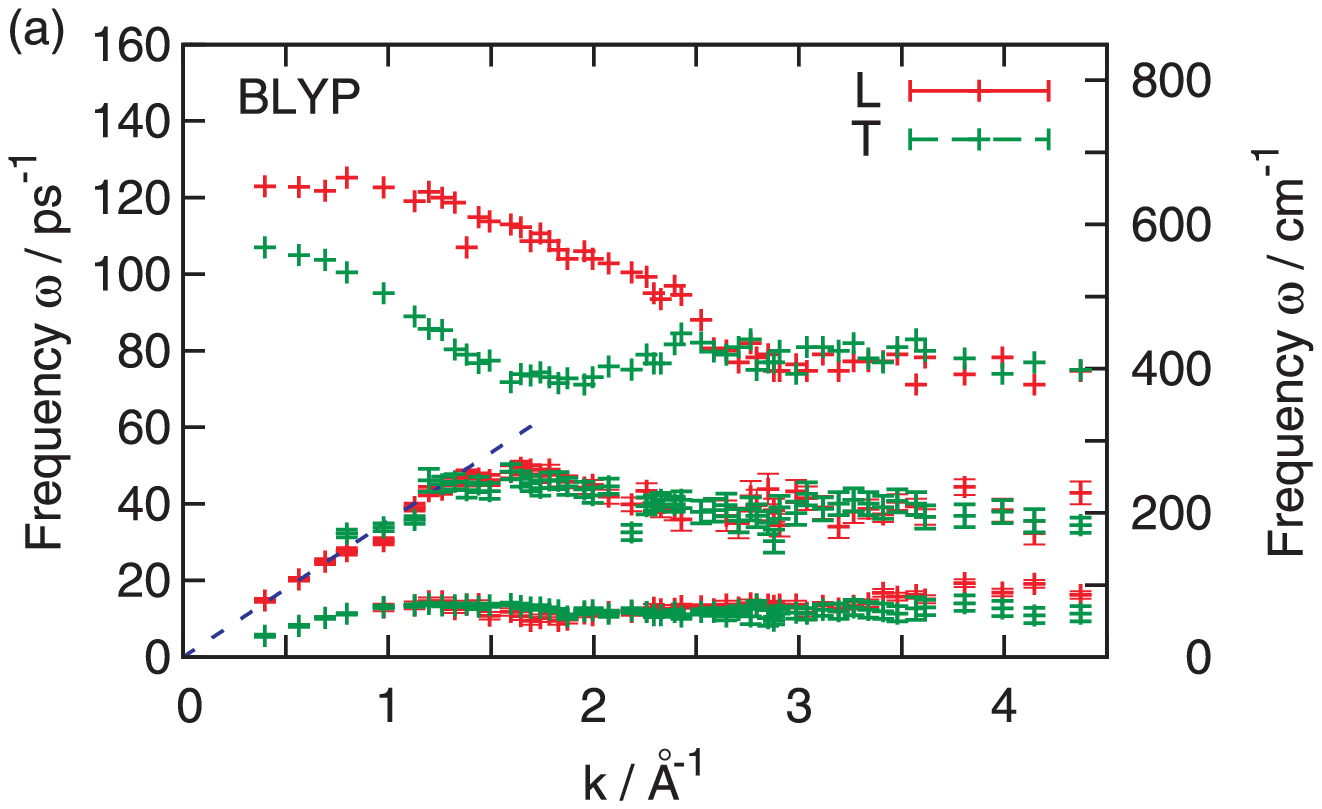}\\
\includegraphics[width=0.65\textwidth]{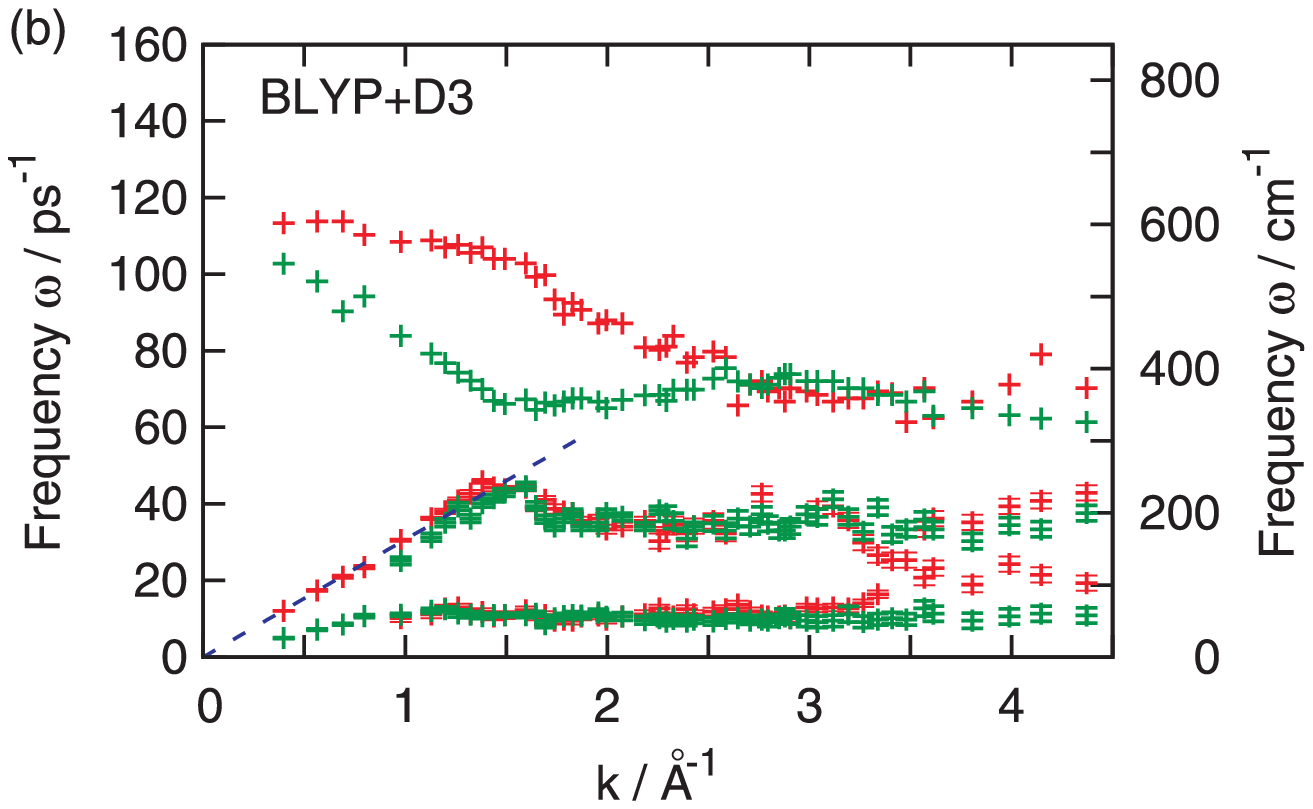}
\end{center}
\vspace{-1mm}
\caption{(Color online) Dispersion of collective excitations obtained from
peak positions of the longitudinal (L) and transverse (T)
current spectral functions $C^\text{L,T}(k,\omega)$ in
 D$_2$O at $T=323.15$~K from BLYP (a) and BLYP+D3 (b) simulations.
The dashed straight line corresponds to a high-frequency linear
dispersion law with a speed of sound of 3640~m/s (a) and 3070~m/s (b).
} \label{disp_D2O}
\end{figure}

The dispersion curves of acoustic and optic collective excitations obtained from
the L and T current spectral functions for D$_2$O with and without D3 dispersion
correction are shown in figure~\ref{disp_D2O}. The small size of the simulated system
does not allow one to observe the transition towards hydrodynamic dispersion law with
adiabatic speed of sound. The dashed straight lines indicate the linear dispersion law with
the apparent high-frequency speed of sound. We obtained a reduction in the value of the apparent
high-frequency speed of sound from 3640 to 3070~m/s when the
D3 dispersion correction was switched on. At the lowest accessible wave number,
the value of the LO-TO gap was also lower when the D3 dispersion correction was applied.
The longitudinal LO frequency was reduced from 123 to 107~ps$^{-1}$, while
the transverse TO frequency changed from 113.8 to 102.8~ps$^{-1}$, resulting
in a reduction of the LO-TO gap from 16 to 11~ps$^{-1}$ due to the effect of
D3 dispersion correction, which transforms the collective dynamics typical of an undercooled state to the regular dynamics of liquid as well as increased the high-frequency dielectric
permittivity [see figure~\ref{strfact}~(b)]. Both simulations with and without D3 dispersion correction
show the same L-T mixing as was reported earlier from classical MD simulations \cite{Sam97}.
In both cases, there is also a similar behaviour in the TO mode, whose frequency decreases
with an increasing wave number up to $k\sim 1.5$~\AA$^{-1}$. At larger wave numbers, it
changes the dependence, showing a small increase up to a wave number $k\sim 2.5$~\AA$^{-1}$, where
it merges with the LO branch. At higher wave numbers, LO and TO branches stay at the same
frequencies. This behaviour of dispersion implies that there is a similar crossover between
the intrinsically collective ($k<1.5$~\AA$^{-1}$) and partial (at $k>1.5$~\AA$^{-1}$)
types of dynamics described in detail in reference~\cite{Bry00a}.

\begin{figure}[!t]
\centerline{
\includegraphics[width=0.65\textwidth]{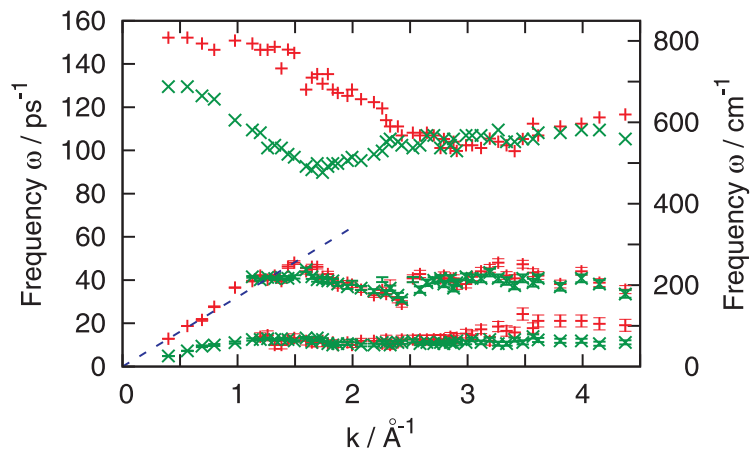}
}
\caption{(Color online) Dispersion of collective excitations obtained from
peak positions of the longitudinal (L) and transverse (T)
current spectral functions $C^\text{L,T}(k,\omega)$ in
 H$_2$O at $T=323.15$~K from BLYP+D3 simulations.
The dashed straight line corresponds to a high-frequency linear
dispersion law with a speed of sound of 3223~m/s.
} \label{disp_H2O}
\end{figure}

In figure~\ref{disp_H2O}, we present the dispersion of longitudinal and transverse collective
excitations in liquid H$_2$O from simulations using BLYP+D3 approach.
The value of apparent high-frequency speed of sound is 3223~m/s, which is in good agreement
with the values $\sim 3100$~m/s obtained from the X-ray scattering experiments on water
close to the melting point~\cite{Mon99} and $\sim 3200$~m/s reported in \cite{Sac04}.
The optic branches of collective excitations
due to the smaller molecular mass are located at higher frequencies than in the case of
heavy water. At the long-wavelength limit, the LO and TO frequencies are 152.2
and 128.2~ps$^{-1}$, respectively, yielding our estimate for the LO-TO gap of 24~ps$^{-1}$.
The frequency of the LO long-wavelength excitation is slightly
lower than the one obtained from classical simulations using a flexible non-polarizable model TIP4P/2005f
\cite{Elt16}. This difference could be due to a softening of the LO mode in AIMD due to the high-frequency
screening effects. Our calculated TO frequency is higher than the one from the classical
simulations~\cite{Elt16}, thus both effects resulting in a smaller long-wavelength LO-TO gap from AIMD.

\section{Conclusions}

We performed {\it ab initio} molecular dynamics simulations for heavy water
with and without the third generation dispersion correction of Grimme and co-workers D3 to
DFT~\cite{Grimme_JCP_2010a}. We found that the D3 dispersion correction strongly
affects the static and dynamic structure of heavy water. Very sensitive to the
thermodynamic state, the behaviour of density-density time correlation
functions showed slow, stretched exponential relaxation, typical of
supercooled liquids, in the case of simulations without D3 dispersion correction,
and an exponential relaxation of regular liquids when the dispersion correction
was used. This finding could be a consequence of a significant reduction of the melting
point of ice in the DFT/BLYP+D3 simulations compared to BLYP-only approach.

Longitudinal and transverse current spectral functions were analyzed in order to estimate
the dispersion of acoustic and optic excitations. The application of the D3 dispersion
correction resulted in a significant reduction of the apparent high-frequency speed
of sound, as well as in a reduction of the gap between long-wavelength
LO and TO excitations. This fact indicates that the D3 dispersion correction leads to higher
values of the high-frequency dielectric permittivity $\varepsilon_{\infty}$. This change is effective, or indirect, since the D3 dispersion correction does not directly involve the electronic structure but only via the changed static and dynamic structure of the water.

Another simulation with D3 dispersion correction was performed for light water at the same
temperature and number density, and it
resulted in an identical static structure as in D$_2$O with the D3 dispersion correction.
Due to a smaller molecular mass, higher frequencies of optic L and T
and of intramolecular modes were obtained. We have a good agreement for
the value of the high-frequency speed of sound with X-ray scattering experiments. The gap
between the long-wavelength LO and TO excitations in light water was obtained as $\sim 24$~ps$^{-1}$, or 127~cm$^{-1}$.
The calculated frequency of the LO excitations from our AIMD is slightly
lower than the one from classical simulations using flexible non-polarizable model TIP4P/2005f
\cite{Elt16}, while frequencies of transverse optic excitations in our {\it ab initio}
simulations with D3 dispersion correction were much higher than the ones from classical simulations.

In all three simulations, we observed the L-T mixing effect, since the L/T current spectral functions
contained an additional peak, corresponding to the T/L excitations, at relatively large wave numbers.
 This L-T mixing effect was observed only for acoustic excitations; the optic modes
appeared at separated frequencies at wave numbers $k<2.5$~\AA$^{-1}$. So far there is no indication
that TO modes can appear in the longitudinal spectra.

Our study gives indications of a significant reduction of the ice I$_h$ melting
temperature in {\it ab initio} simulations with D3 dispersion correction applied. There is thus a quest for large
AIMD simulations with D3 dispersion correction on the melting of ice I$_h$.


\section*{Acknowledgements}

APS acknowledges Guillaume Ferlat, A Marco Saitta and Rodolphe Vuilleumier for many illuminating discussions on simulations of water, and the computing resources at the Universit\"{a}t Z\"{u}rich, Zurich, and Centro Svizzero di Calcolo Scientifico, CSCS, Ticino, Switzerland.


%

\clearpage

\ukrainianpart

\title{Дослідження колективних збуджень у воді методом першопринципної
молекулярної динаміки:  ефект \\ ван~дер~вальсівських поправок на дисперсію \\ колективних
збуджень}
\author{Т. Брик\refaddr{label1,label2}, А.П. Сейтсонен\refaddr{label3,label4}
}
\addresses{
\addr{label1} Інститут фізики конденсованих систем НАН України, вул. І. Свєнціцького, 1, 79011 Львів, Україна
\addr{label2} Інститут прикладної математики та фундаментальних наук,
Національний Університет ''Львівська Політехніка'', 79013 Львів, Україна
\addr{label3} Інститут хімії, Університет Цюріха, Вінтертурерштрассе, 190,  CH-8057 Цюріх, Швейцарія
\addr{label4} Хімічний факультет, Вища нормальна школа, 24 вул. Льомон, F-75005 Париж, Франція
}

\makeukrtitle

\begin{abstract}
\tolerance=3000%
Колективна динаміка у рідкій воді активно досліджується експериментальними, теоретичними
 методами та комп'ютерними симуляціями.
Представлено моделювання методом першопринципної молекулярної динаміки для важкої та
звичайної води при температурі 323.15~K, чи 50$^\circ$C. Моделювання для важкої води були проведені
з та без дисперсійних поправок. Ми отримали, що дисперсійні поправки (DFT-D3) суттєво
змінюють релаксацію часових кореляційних функцій густина-густина з повільної, яка є
типовою для переохолодженого стану, до експоненційного спаду як для звичайних рідин.
Це означає суттєве зменшення точки плавлення льоду для моделювання з DFT-D3. Аналіз
повздовжних (L) та поперечних (T) спектральних функцій потоків дозволив визначити
дисперсії акустичних та оптичних збуджень та спостерігати ефект  L-T змішування.
Дисперсійні поправки D3 зсувають до нижчих частот L та T оптичні (O) моди та приводять
до меншої на майже тридцять процентів щілини між LO та TO довгохвильовими збудженнями,
що може бути наслідком більшої високочастотної діелектричної проникності в моделюванні
з дисперсійними поправками. Моделювання для звичайної води з дисперсійними поправками
дає вищі частоти оптичних збуджень, ніж для D$_2$O, та
щілину LO-TO порядку 24~пс$^{-1}$ (127~см$^{-1}$) у довгохвильовій границі.

\keywords колективні збудження, оптичні моди, вода, важка вода, ван~дер~вальсівські поправки,
першопринципне комп'ютерне моделювання
\end{abstract}

\end{document}